\documentclass[prb,twocolumn]{revtex4}
\usepackage{times}
\usepackage{wrapfig}
\usepackage{amsfonts}
\usepackage{amssymb}
\usepackage{psfrag}
\usepackage{graphicx}
\usepackage{amsmath}

\usepackage{color}

\newcommand{\dd}{\mbox{d}}

\begin{document} 

\title{On the Ground State of Electron Gases at Negative
Compressibility}

\author{Adriaan M. J. Schakel}

\affiliation{National Chiao Tung University, Department of
Electrophysics, Hsinchu, 30050, Taiwan, R.O.C.}

\date{\today} 

\begin{abstract} 
Two- and three-dimensional electron gases with a uniform neutralizing
background are studied at negative compressibility.  Parametrized
expressions for the dielectric function are used to access this
strong-coupling regime, where the screened Coulomb potential becomes
overall attractive for like charges.  Closely examining these
expressions reveals that the ground state with a periodic modulation of
the charge density, albeit exponentially damped, replaces the
homogeneous one at positive compressibility.  The wavevector
characterizing the new ground state depends on the density and is
complex, having a positive imaginary part, as does the homogeneous
ground state, and real part, as does the genuine charge density wave.
\end{abstract}

\maketitle
\section{Introduction}
The nature of the metallic state of a dilute two-dimensional electron
gas (2DEG) in high-mobility silicon MOSFETs, first observed by
Kravchenko {\it et al.}  \cite{Kravchenko} nearly a decade ago, has
still not been established.  Various scenarios have been proposed to
explain the new conducting state, ranging from superconducting, to
non-Fermi liquid, to percolation, to classical (see Refs.\
\cite{P^2,AKS} for overviews and references to the original literature).
The unique character of this state is reflected by its thermodynamic
signatures of both negative pressure and compressibility \cite{EPW}.

Although recognized as a theoretical possibility for quite some time---at
least in three dimensions \cite{PN}, an electron gas with a uniform
neutralizing background was expected to be unstable at negative
compressibility, and the model meaningless.  Later theoretical
\cite{Ichimaru} and also numerical \cite{TC} studies nevertheless
considered both the region with positive and negative compressibility
without encountering conceptual difficulties.  

Eisenstein, Pfeiffer, and West \cite{EPW} more recently concluded that
the global features of their 2DEG compressibility data to very low
densities, where the system appears to undergo a quantum phase
transition to an insulating state \cite{AKS}, could be explained simply
using the Hartree-Fock approximation, which consists of including the
exchange contribution to the ground-state energy of a free fermion gas.
Their analysis does not account for impurities, which are of paramount
importance at and beyond the metal-insulating (MI) transition.

Ilani {\it et al.}  \cite{IYMS} reached similar conclusions based on
their compressibility data of a dilute two-dimensional hole gas (2DHG).
By placing several single electron transistors directly above the 2DHG,
they could in addition determine possible spatial variations in the
system.  They concluded that the metallic state at negative
compressibility is homogeneous in space and well described by the
Hartree-Fock approximation.  As the system crosses to the insulating
state and the Hartree-Fock approximation ceases to be applicable, they
found the system to become spatially inhomogeneous.  They interpreted
this observation as supporting scenarios in which the MI transition is
described as a percolation process \cite{Efros,HeXie,Meir}

Using density functional theory in the local density approximation, Shi
and Xie \cite{ShiXie} recently investigated the spatial distribution of
the electron number density of a 2DEG in the presence of impurities.  As
usual, impurities were included by coupling the particle number density
to a fluctuating potential with a random distribution.  The parameter
characterizing the Gaussian distribution, which we will refer to as the
impurity strength, determines the roughness of the impurity landscape.
Their numerical study incorporates the Monte Carlo data on the
ground-state energy of a clean system by Tanatar and Ceperly
\cite{TC}. For given impurity strength, the study shows that at some
average density $n$ islands of very low densities form in a metallic sea
of high densities.  At higher average densities, the sea level is high
enough to fill all of the valleys of the impurity landscape, and the
system is homogeneous.  With decreasing $n$, the sea level drops and the
insulating islands grow.  At a critical value, the islands percolate the
system, which at this percolation threshold becomes insulating.  The sea
level now dropped to the extent that the electrons are confined to the
valleys of the impurity landscape.  According to this picture, the
transition to the insulating state takes place at lower densities for
cleaner systems.  This is in accord with experiments, where the lowest
critical density, $n_{\rm c} = 7.7\pm 0.4\times 10^9 \mbox{cm}^{-2}$,
was observed in an exceptionally clean 2DHS \cite{Yoonetal}.  Shi and
Xie \cite{ShiXie} also studied the importance of the Coulomb interaction
by comparing with the case where this interaction is turned off.  They
found that at an average density where the interacting system was still
metallic, the free electron gas, which is insulating, had formed a few
isolated lakes of high density at valleys of the impurity landscape.

In this paper, we postulate on the nature of the ground state of
electron gases at negative compressibility.  In doing so, we ignore
impurities, as justified by the experimental findings that this metallic
state is homogeneous in space and that the compressibility data is
already well described by the Hartree-Fock approximation of a clean
electron gas.  The study emphasizes the three-dimensional electron gas
(3DEG) as much more information from theoretical studies is available
than for the 2DEG.  Most of the 3D conclusions, however, also apply to
the 2D case discussed in Sec.\ \ref{sec:2DEG}.  The following two
sections focus on the changes brought about when going from positive to
negative compressibility.  One of the most surprising changes is that a
test particle acquires a screening cloud in its immediate vicinity which
overcompensates the test charge, so that the Coulomb interaction becomes
{\it attractive} for like charges.  The changes are also discussed from
the perspective of Fermi liquid theory, describing the electron gas after
the screening mechanism has taken effect and resulted in only
short-range interactions.  The main result of this paper is contained in
Sec.\ \ref{sec:CDW}, where it is argued that the ground state of an
electron gas at negative compressibility is a charge density wave (CDW),
although exponentially damped.  The phenomena considered here are by no
means specific to an electron gas, but also appear in a charged Bose gas
as discussed in Sec.\ \ref{sec:Bose}.
\section{Screening vs. Overscreening}
A 3DEG is characterized by the screening of charges.  When a test
particle of charge $Z$ is placed at the origin, $\rho(x) = Z
\delta({\bf x})$, the system responds by rearranging its charge
distribution to screen the external charge.  The charge density
$\rho_{\rm ind}$ thus induced,
\begin{equation} 
\rho_{\rm ind}(x) = Z \int \frac{\dd^3q}{(2 \pi)^3} \left[
\frac{1}{\epsilon(q,0)} - 1 \right] {\rm e}^{{\rm i} {\bf q} \cdot {\bf
x}},
\end{equation} 
is determined by the dielectric function $\epsilon(q,0)$ at zero
frequency, which encodes the static screening effects of the 3DEG.  The
total induced charge 
\begin{equation} 
Q_{\rm ind} = \int \dd^3 x \, \rho_{\rm ind}(x) = \rho_{\rm ind}(q=0) =
-Z
\end{equation} 
cancels the charge of the test particle since the screening factor
$1/\epsilon(q,0)$ vanishes at zero wavevector $q=|{\bf q}|$.  The test
charge is therefore always perfectly screened.  However, the induced
charge density differs in how it is distributed, depending on the
electron number density $n$, or, equivalently, the ratio $r_{\rm s}=
a/a_0$ of the average interparticle distance $a = (3/4 \pi n)^{1/3}$ to
the Bohr radius $a_0 = \hbar^2/m e^2$.  (For valence electrons in 3D
metals, $r_{\rm s}$, characterizing the strength of the Coulomb
interaction, ranges from \cite{PN} $1.8$ to $5.6$.)

At weak coupling ($r_{\rm s} < 1$), corresponding to high electron
number densities, a screening cloud of opposite charge surrounds the
test charge, and the screened Coulomb potential decreases exponentially
with increasing distance.  On its tail, far away from the test charge,
the potential has superimposed a small oscillatory modulation of
wavevector $q=2 k_{\rm F}$, with $k_{\rm F}$ the Fermi wavevector.
These Friedel oscillations, leading to a periodic change in sign of the
Coulomb potential, originate from a singularity in the dielectric
function at $q=2 k_{\rm F}$, where electron-hole excitations start to
develop an energy gap \cite{PN}.

A 3DEG's response to a test charge fundamentally changes at larger
values of the coupling constant $r_{\rm s}$ because of a qualitative
change in the dielectric function.  Namely, at some value $r_{\rm s} =
\bar{r}_{\rm s}$, the dielectric function becomes negative for small
wavevectors \cite{Gold}.  Rather than surrounded by a screening cloud of
opposite charge in its immediate vicinity, the test charge now becomes
overscreened.  The Coulomb potential rapidly drops below zero with
increasing distance and becomes attractive for like charges. Further
away from the test charge, the potential exhibits an exponentially
damped oscillatory behavior and periodically changes sign similar to
Friedel oscillations.  The initial drop of the Coulomb potential to
negative values and the resulting overscreening is, however, unrelated
to Friedel oscillations \cite{CaGo}.  Overscreening in a plasma was
first noted in a 2D Bose gas with a $1/x$ Coulomb potential \cite{HiFr}.

The fundamental change in the screening behavior of a 3DEG is paralleled
by a modification of its ground state.  At the critical electron number
density, where the dielectric function becomes negative for small
wavevectors and a test charge overscreened, the compressibility changes
sign as well.  As argued in the following, the homogeneous ground state of
the regime with positive compressibility then gives way to a ground
state with a periodic modulation of the charge density, which is,
however, exponentially damped.

\section{Negative Compressibility}
At weak coupling, the dielectric function can be written for small
wavevectors as \cite{PN}
\begin{equation}   \label{epsilonsmall}
\lim_{q \to 0} \epsilon(q,0) = 1 - v(q) \chi_{\rm sc}(0,0),
\end{equation}  
with $v(q) = 4 \pi e^2/q^2$ the Fourier transform of the (unscreened)
Coulomb potential, and $\chi_{\rm sc}(0,0)$ the {\it screened}
density-density response function at zero wavevector and frequency.
The screened response function, a theoretic construct to be
distinguished from the physical one, measures the response to a screened
external field rather that the external field itself.  Physically, this
function describes a fictitious system with only short-range
interactions that are remnants of the long-range Coulomb interaction
after the screening mechanism of the 3DEG has taken effect \cite{PN}.
Derived from the Coulomb interaction, the short-range interactions of
the fictitious system vanish in the limit $r_{\rm s} \to 0$.

In writing Eq.\ (\ref{epsilonsmall}), the ${\cal O}(q^2)$ term in the
Taylor expansion of $\chi_{\rm sc}(q,0)$ is assumed to be substantially
smaller than $1$.  This is true only at weak coupling.  For example,  when
$r_{\rm s} =1$, the value of the constant term in Eq.\
(\ref{epsilonsmall}) is reduced already by about 6~\%.

The compressibility sum rule relates the screened response function to
the compressibility $\kappa$ of the 3DEG via \cite{PN}
\begin{equation} \label{compressibility}
\lim_{q \to 0} \chi_{\rm sc}(q,0) = - \frac{\partial n}{\partial \mu} =
- n^2 \kappa,
\end{equation}
with $\mu$ the chemical potential.  If positive, the compressibility can
be expressed in terms of the speed of sound $c$ in the fictitious system
with only short-range interactions as 
\begin{equation} 
n \kappa = 1/m c^2.
\end{equation}   
Equation (\ref{epsilonsmall}) can then be cast in the equivalent form
\begin{equation} \label{epsilonsmall'}
\lim_{q \to 0} \epsilon(q,0) = 1 + \frac{\omega_{\rm pl}^2}{c^2 q^2},
\end{equation} 
corresponding to the spectrum $\omega^2 = \omega_{\rm pl}^2 + c^2 q^2$
of the plasma mode, where $\omega_{\rm pl}$ denotes the plasma
frequency, $\omega^2_{\rm pl} = 4 \pi n e^2/m$.  The plasmon spectrum
differs from the gapless spectrum $\omega^2 = c^2 q^2$ of the sound mode
of the fictitious system in that, due to the long range of the Coulomb
interaction, the former has an energy gap.  At short wavelengths $(c q
> \omega_{\rm pl})$, the difference is negligible, thereby allowing use
of the plasmon to access, via $c$, the screened interaction as a
function of the coupling constant---at least for $r_s <1$.

In the limit $r_{\rm s} \to 0$, the screened response function reduces
to (minus) the density of states $\nu_0 = m \hbar k_{\rm F}/\pi^2$ at
the Fermi surface.  Simultaneously, $c^2 \to c_0^2 = v_{\rm F}^2/3$,
with $v_{\rm F}= \hbar k_{\rm F}/m$ the Fermi velocity, and the
Thomas-Fermi approximation for the dielectric function is recovered.

Owing to its short-range interactions, the fictitious system can be
described by Fermi liquid theory \cite{PN}.  The elementary excitations
are fermionic quasiparticles of mass $m^*$, with an interaction
characterized by spin symmetric (s) and spin antisymmetric (a) Landau
parameters $F^{\rm s,a}_\ell$, where $\ell$ denotes the angular momentum
channel.  The Landau parameters depend on $r_{\rm s}$ and vanish in the
limit $r_{\rm s} \to 0$, where also $m^* \to m$.  The speed of sound in
the fictitious system can be expressed in these parameters as follows
\begin{equation} 
c^2 = (1 + F^{\rm s}_0) (m/m^*) c_0^2.
\end{equation}   
Due to the Pauli exclusion principle, even with an attractive
interaction, a Fermi liquid can still support a sound mode, provided
that $F_0^{\rm s}>-1$.

Approximate calculations \cite{Rice,PN} indicate that the effective mass
is comparable to the free electron mass, while the Landau parameter
$F_0^{\rm s}$ is negative.  The latter implies an {\it attractive}
quasiparticle interaction in the spin-symmetric, $\ell=0$ channel,
although it derives from the Coulomb interaction, which repulses like
charges.  This unexpected finding has the same origin as the
overscreening of a test charge.

At the level of the first-order correction to the ground-state energy of
a free Fermi gas, which itself is a 1-loop result, the origin can be
understood as follows \cite{PN}.  Two Feynman diagrams contribute to the
2-loop, or Hartree-Fock correction: the direct, or Hartree term
containing two fermion loops, and the exchange term containing only one
fermion loop.  Owing to overall charge neutrality, the direct term does
not contribute to the ground-state energy, so that only the exchange
term remains.  Containing an odd number of fermion loops, this term
comes with a minus sign, thus reversing the sign of the Coulomb
interaction.  Consequently, the ground-state energy per electron and
also the pressure decrease with increasing coupling constant for $r_{\rm
s} < 1$, eventually becoming negative.

The inverse compressibility of the 3DEG and therefore the speed of sound
$c$ in the fictitious system also decrease with increasing $r_{\rm s}$.
At a certain value $r_{\rm s} = \bar{r}_{\rm s}$, with $\bar{r}_{\rm s}
\approx 5.25$ according to estimates \cite{Ichimaru}, the inverse
compressibility becomes negative and the speed of sound drops to zero,
implying that the factor $(1 + F^{\rm s}_0)/m^*$ should vanish.  The
approximate calculations \cite{Rice,PN} indicate that the quasiparticle
mass increases only slightly with increasing $r_{\rm s}$, while the
values of $F^{\rm s}_0$ for $r_{\rm s} =2,3,4$ show a tendency towards
$-1$ around $r_{\rm s} = \bar{r}_{\rm s}$.  We conjecture that precisely
at this point, $F^{\rm s}_0=-1$.  In Fermi liquid theory \cite{PN}, this
value of the Landau parameter $F^{\rm s}_0$, where $\chi_{\rm sc}(0,0)$
diverges, signifies the onset of instability, with the homogeneous ground
state becoming unstable towards density fluctuations.
\section{Exponentially Damped CDW}  \label{sec:CDW}
The vanishing of the sound mode of the fictitious system affects the
plasmon spectrum.  The general condition for plasma oscillations at a
frequency $\omega$ is \cite{PN} $\epsilon (q,\omega)=0$.  At zero
frequency, or energy, the condition reduces to
\begin{equation} \label{condition}
q^2 \epsilon (q,0)=0,
\end{equation}  
where an additional factor $q^2$ is included for convenience.  A
physical solution $q(r_{\rm s})$ of this condition with a positive real
part denotes a time-independent, i.e., frozen-in modulation of the
charge density.

Consider condition (\ref{condition}) first at weak coupling, where the
dielectric function reduces to the form (\ref{epsilonsmall'}) in the
limit of long wavelengths.  This gives $q^2(r_{\rm s}) = - \omega_{\rm
pl}^2/c^2$, leading to a {\it purely} imaginary wavevector as a
solution, and a screening length $\lambda = c/\omega_{\rm pl}$.  In the
limit $r_{\rm s} \to 0$, this formula reduces to the Thomas-Fermi result
\begin{equation} \label{screeningf}
\lambda/a = (\pi \alpha/ 4 r_{\rm s} )^{1/2}, \;\;\;\; \alpha = (4/9
\pi)^{1/3}.
\end{equation} 
Being proportional to the inverse square root of the coupling constant,
the screening length (\ref{screeningf}) measured in units of the
interparticle distance $a$ becomes infinite as $r_{\rm s}$ approaches
zero.

At $r_{\rm s} = \bar{r}_{\rm s}$, where the dielectric function becomes
negative for small wavevectors, resulting in the overscreening of a test
charge, the solution of the condition (\ref{condition}) changes
qualitatively.  In the region $r_{\rm s} > \bar{r}_{\rm s}$, the plasmon
spectrum initially decreases with increasing wavevector, until reaching
a (positive) minimum after which it increases \cite{STLS}.

The unique character of the point $r_{\rm s} = \bar{r}_{\rm s}$ can also
be noted when considering the spatial average of the electrostatic
potential generated by the static test charge at the origin
\begin{equation} \label{average}
\int \dd^3 x \, \varphi(x) = \varphi(q=0) = \frac{Z}{e^2 n^2 \kappa},
\end{equation} 
where $\varphi (q) = 4 \pi Z/\epsilon(q,0) q^2$.  When the inverse
compressibility becomes negative, the overall potential changes sign
as well.

To investigate the strong-coupling regime, we use a parametrized
expression for the local-field correction $G(q)$ as it appears in the
generalized random phase approximation of the screened response
function:
\begin{equation} \label{chisc} 
\chi_{\rm sc}(q,0) = \frac{\chi_0(q,0)}{1 + v(q) G(q) \chi_0(q,0)}
\end{equation} 
proposed by Ichimaru and Utsumi \cite{IcUt}.  Here, $\chi_0(q,0)$
denotes the response function of a free Fermi gas at zero frequency.
The random phase approximation is recovered by setting $G(q)$ to unity.
Since only the zero-frequency response function is required to study the
condition (\ref{condition}), we can sidestep the difficulties which
arise when the frequency dependence is included in $G(q)$ to arrive at
the dynamic correction.  The parametrized expression, which applies to
the range $0 < r_{\rm s} < 15$, incorporates Monte Carlo data on the
ground-state energy \cite{CeAd} as well as the ladder diagram
calculation of the pair distribution function at zero separation
\cite{Yashuhara}.  Importantly, the resulting dielectric function
satisfies a number of exact boundary conditions and sum rules, including
the compressibility sum rule (\ref{compressibility}).  As noted in the
overview \cite{Gorobchenko}, these features and its simplicity makes the
parametrized expression proposed in Ref.\ \cite{IcUt} convenient for
applications.

The compressibility sum rule determines the behavior of the local-field
correction at long wavelengths.  With the definition
\begin{equation} 
\lim_{q \to 0} G(q) = \gamma_0(r_{\rm s}) \hat{q}^2,
\end{equation} 
where $\hat{q}=q/k_{\rm F}$, it follows from Eq.\
(\ref{compressibility}) and a similar expression for the noninteracting
system with compressibility $\kappa_0$, that the coefficient
$\gamma_0(r_{\rm s})$ is related to the compressibility via
\begin{equation} 
\frac{\kappa_0}{\kappa} = 1 - \frac{4 \alpha}{\pi} \gamma_0(r_{\rm s})
r_{\rm s}.
\end{equation} 
The compressibility of a 3DEG can be extracted from the Monte Carlo data
of Ceperly and Alder \cite{CeAd} thus fixing $\gamma_0(r_{\rm s})$.  In
particular, when the inverse compressibility changes sign it becomes
\begin{equation} 
\gamma_0(\bar{r}_{\rm s}) = \frac{\pi}{4 \alpha} \frac{1}{\bar{r}_{\rm
s}}.
\end{equation} 
Because the compressibility sum rule is satisfied, the parametrized
expression for the dielectric function,
\begin{equation} \label{para}
\epsilon(q,0) = 1 - v(q) \chi_{\rm sc}(q,0),
\end{equation} 
with $\chi_{\rm sc}$ given by Eq.\ (\ref{chisc}), becomes negative for
small wavevectors also at $r_{\rm s} = \bar{r}_{\rm s}$, as it should.

With the expression (\ref{para}) substituted and the left hand expanded
in a Taylor series to order $q^4$, Eq.\ (\ref{condition}) leads to the
condition:
\begin{equation} \label{expand}
a_0 + a_2 \hat{q}^2 + a_4 \hat{q}^4 = 0,
\end{equation}  
valid at long wavelengths.  The quartic term is included because the
first term in the expansion,
\begin{equation} 
a_0 = \frac{(4 \alpha/\pi) r_{\rm s}}{1 - (4 \alpha/\pi) \gamma_0(r_{\rm
s})r_{\rm s}},
\end{equation} 
changes sign at $r_{\rm s} = \bar{r}_{\rm s}$.  The coefficients $a_2$
and $a_4$ have no simple analytic representation, depending on the
specific parametrization of the local-field correction $G(q)$.  They are
best represented simply by their numerical values for each $r_{\rm s}$.
The coefficients diverge at $r_{\rm s} = \bar{r}_{\rm s}$.  As a result
of which, a small region just below $\bar{r}_{\rm s}$ is numerically
inaccessible.
\begin{figure}
\begin{center}
\psfrag{im}[t][t][.8][-10]{Im[$\hat{q}(r_{\rm s})$]}
\psfrag{re}[t][t][.8][15]{Re[$\hat{q}(r_{\rm s})$]}
\psfrag{rs}[t][t][.8][0]{$r_s$}
\psfrag{q}[t][t][.8][0]{}
\psfrag{1}[t][t][.8][0]{$1$}
\psfrag{2}[t][t][.8][0]{$2$}
\psfrag{4}[t][t][.8][0]{$4$}
\psfrag{6}[t][t][.8][0]{$6$}
\psfrag{8}[t][t][.8][0]{$8$}
\psfrag{10}[t][t][.8][0]{$10$}
\psfrag{12}[t][t][.8][0]{$12$}
\psfrag{14}[t][t][.8][0]{$14$}
\psfrag{0.5}[t][t][.8][0]{$0.5$}
\psfrag{1.5}[t][t][.8][0]{$1.5$}
\includegraphics[width=8.cm]{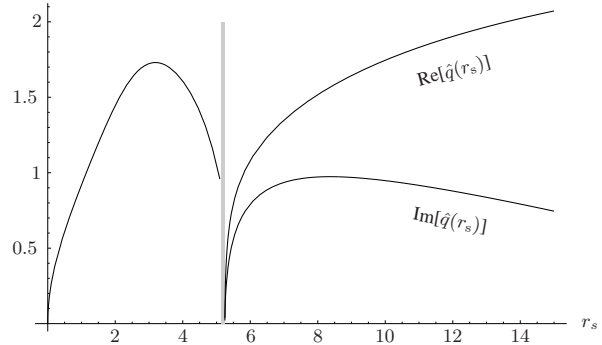}
\end{center}
\caption{Real (Re) and imaginary (Im) part of the wavevector
$\hat{q}(r_{\rm s})$ solving the condition (\ref{expand}).  The grey
region just below $r_{\rm s} = \bar{r}_{\rm s} \approx 5.25$ could not
be accessed because of numerical instabilities in the expansion
coefficients. \label{fig:q_sol}}
\end{figure}

The physical solution $q(r_{\rm s})$ of the condition (\ref{expand}),
which is a quadratic equation in $q^2$, remains purely imaginary in the
entire regime where $1/\kappa >0$, as it is at weak coupling (see Fig.\
\ref{fig:q_sol}).  The corresponding screening length $\lambda = {\rm
i}/q(r_{\rm s})$, being infinite at $r_{\rm s}=0$, decreases with
increasing coupling constant until it reaches a minimum $\lambda/a
\approx 0.30$ at $r_{\rm s} \approx 3.2$.  Further increasing $r_{\rm
s}$ surprisingly increases $\lambda/a$ again until it becomes infinite
once more at $r_{\rm s} = \bar{r}_{\rm s}\approx 5.25$, where the
inverse compressibility changes sign and the sound mode of the
fictitious system vanishes.  The wavevector solving the condition at
long wavelengths is zero and the Coulomb interaction unscreened, as it
was at $r_{\rm s} = 0$, according to this definition of the screening
length.

Remarkably, plotting the inverse dielectric function at this value of
$r_{\rm s}$, indicates, to within numerical accuracy, no dispersion for
wavevectors smaller than the Fermi wavevector and $1/\epsilon(q,0)=0$
in this range $0 \leq q \leq k_{\rm F}$.

For $r_{\rm s} > \bar{r}_{\rm s}$, the physical solution $q(r_{\rm s})$
becomes genuinely complex, having an imaginary and a positive real part
as well.  The latter implies the onset of instability, where the
homogeneous ground state becomes unstable towards a periodic spatial
modulation of the charge density.  Owing to the imaginary part,
fluctuations are, however, still exponentially screened.  The ground
state at negative compressibility is referred to in this paper as an
{\it exponentially damped CDW} since it combines a homogeneous ground
state with exponential screening and a CDW.  The real part of the
wavelength of the exponentially damped CDW is infinite at $r_{\rm s} =
\bar{r}_{\rm s}$ and decreases with increasing $r_{\rm s}$ in the
remaining range where the parametrized expression for $\epsilon(q,0)$ on
which our analysis is based applies, i.e., $r_{\rm s}<15$.  The
screening length $\lambda$ also starts at infinity and initially
decreases with increasing $r_{\rm s}$.  But, after reaching a minimum
$\lambda/a \approx 0.54$ at $r_{\rm s} \approx 8.37$, it increases
again.

Apart from being exponentially damped, a CDW in a 3DEG differs from a
CDW appearing in solids also in that its wavevector is not necessarily
given by $q=2 k_{\rm F}$, but varies with $r_{\rm s}$ (see Fig.\
\ref{fig:q_sol}).  In a solid, the CDW arises due to electron-phonon
interactions, and the charge modulation is accompanied by a periodic
lattice distortion both of wavevector $q=2 k_{\rm F}$ \cite{Gruener}.

The screening length $\lambda$ determined by the imaginary part of the
wavevector solving condition (\ref{expand}) does not monotonically
decrease with increasing coupling constant.  Since the screening
mechanism is expected to become more effective at stronger coupling to
minimize the effect of the increasing Coulomb interaction, $\lambda$
fails to provide a proper measure for this mechanism.  A more relevant
length scale is the {\it short-range} screening length $\lambda_{\rm
s}$, in which the deviation of the screened from the unscreened Coulomb
potential is measured at short distances from a test charge
\cite{Ichimaru_book}.  It is defined by writing the electrostatic
potential generated by a static test charge at the origin [see below
Eq.\ (\ref{average})] as
\begin{equation} 
\varphi(x) = \frac{Z}{x} \left[ 1 + \frac{2}{\pi} \int_0^\infty \dd q
\left( \frac{1}{\epsilon(q,0)} -1 \right) \frac{\sin(qx)}{q} \right]
\end{equation} 
and expanding the right side in a Taylor series for small $x$,
\begin{equation} 
\varphi(x) = \frac{Z}{x} \left( 1 - \frac{x}{\lambda_{\rm s}} + \cdots
\right),
\end{equation} 
with
\begin{equation} 
\frac{a}{\lambda_{\rm s}} = \left(\frac{18}{\pi^2}\right)^{1/3}
\int_0^\infty \dd \hat{q} \left[1 - \frac{1}{\epsilon(\hat{q},0)}
\right].
\end{equation} 
Since $1/\epsilon(\hat{q},0) < 1$, $\lambda_{\rm s}$ is always positive.
The short-range screening length coincides with the Thomas-Fermi
screening length in the limit $r_{\rm s} \to 0$.  As expected for a
system coping with an interaction that becomes increasingly stronger,
$\lambda_{\rm s}$ monotonically decreases (roughly as $r_{\rm
s}^{-1/2}$) with increasing coupling constant in the entire range $0 <
r_{\rm s} < 15$ where the parametrized expression for $\epsilon(q,0)$
applies.  The ratio $\lambda_{\rm s}/a$ is unity around $r_{\rm s} =
0.78$.  Whereas at $r_{\rm s} = \bar{r}_{\rm s}$, its value is reduced
to $\lambda_{\rm s}/a \approx 0.39$.

The exponentially damped CDW in a 3DEG arises in the nonperturbative,
strong-coupling regime.  It is worth considering an example where a
similar ground state arises at weak coupling, to access it in
perturbation theory.  Such an example is provided by a charged Bose gas.
\section{Charged Bose Gas}  \label{sec:Bose}
A {\it free} Bose and Fermi gas differ in that, owing to the Pauli
exclusion principle, the latter can support a sound mode at zero
temperature, whereas the former cannot.  The single-particle excitation
with the spectrum $\omega = \hbar q^2/2m$ is therefore the only gapless
mode available.  By repeating the argument leading to the plasmon
spectrum of a 3DEG at weak coupling, we obtain the spectrum $\omega^2 =
\omega_{\rm pl}^2 + \hbar^2 q^4/4m^2$ of a charged Bose gas in the limit
$r_{\rm s} \to 0$, with the same expressions for the plasma frequency,
$\omega^2_{\rm pl} = 4 \pi n e^2/m$, and $r_{\rm s}$ as for a 3DEG.  The
plasma spectrum of the charged Bose gas at weak coupling corresponds to
the dielectric function
\begin{equation} \label{epsilonb}
\lim_{q \to 0} \epsilon(q,0) = 1 + \frac{\omega_{\rm pl}^2}{\hbar^2
q^4/4 m^2}.
\end{equation} 
These formulas agree with the perturbative results first obtained in
Ref.\ \cite{Foldy} and Ref.\ \cite{HoFr}, respectively, using
Bogoliubov's method.

A solution of the condition (\ref{condition}) with the dielectric
function (\ref{epsilonb}) is given by 
\begin{equation} 
q(r_{\rm s})=(1+{\rm i}) (m \omega_{\rm pl}/\hbar)^{1/2}.
\end{equation}   
The resulting ground state is an exponentially damped CDW with both the
wavelength and screening length expressed as
\begin{equation} 
\lambda/a = 1/(3 r_{\rm s})^{1/4},
\end{equation}  
cf.\ the analogous expression (\ref{screeningf}) for a 3DEG.

As for a 3DEG, the exponentially damped CDW of a charged Bose gas in 3D
has a negative compressibility as well. Specifically, the energy per
particle given by \cite{Foldy}
\begin{equation} 
\frac{E}{N} = -A \frac{e^2}{2 a_0} \frac{1}{r_{\rm s}^{3/4}}
\end{equation} 
to the lowest order in the loop expansion leads to
\begin{equation} 
\frac{1}{\kappa} = - \frac{5 A}{16} \frac{e^2 }{2 a_0} \frac{n}{r_{\rm
s}^{3/4}} ,
\end{equation} 
with $A \approx 0.8031$.  A negative compressibility is possibly a
generic characteristic of an exponentially damped CDW.
\section{2DEG}  \label{sec:2DEG}
Next, 2DEGs with a $1/x$ Coulomb potential and average interparticle
distance $a = (\pi/n)^{1/2}$ are considered.  The shift to 2D is
facilitated by replacing the 3D Fourier transform $4 \pi e^2/q^2$ of the
$1/x$ potential with $2 \pi e^2/q$ in the above equations.  Doing so
leads to a plasma frequency
\begin{equation} \label{2dplas}
\omega^2_{\rm pl} = 2 \pi n e^2q/m,
\end{equation} 
which depends on $q$ and tends to zero for vanishing wavevectors.  That
is, although harder than the gapless modes of the corresponding
fictitious fermionic and bosonic systems, the resulting plasma modes
are, unlike their 3D counterparts, gapless.  Most of the striking
features of the 3D systems, such as negative compressibility \cite{TC},
the Landau parameter $F^{\rm s}_0$ taking the value $-1$ at the point
where the inverse compressibility changes sign \cite{KCM}, and
overscreening \cite{HiFr,Gold} are nevertheless also found in 2D.

One noteworthy difference is that the condition (\ref{condition}) in 2D
with the weak-coupling expression for the dielectric function
(\ref{epsilonsmall'}) and the plasma frequency (\ref{2dplas}),
\begin{equation}  \label{2dcon}
q^2 + 2 \pi n e^2q/m c^2 = 0,
\end{equation} 
has no physical solution.  In accordance with the gapless plasmon
spectrum, this implies the absence of Thomas-Fermi screening.

The 2DEG's dielectric function $\epsilon(q,0)$ has not been determined
to the extent the 3DEG's function has.  However, the local-field
correction proposed in Ref.\ \cite{Iwamoto}, incorporating the
variational Monte Carlo data on the ground-state energy and the
compressibility by Tanatar and Ceperley \cite{TC}, serves our purposes
as the corresponding dielectric function satisfies the compressibility
sum rule.  Specifically, with the definition 
\begin{equation} 
\lim_{q \to 0} G(q) = \gamma_0(r_{\rm s}) \hat{q}
\end{equation} 
appropriate for 2D, it follows from the compressibility sum rule
(\ref{compressibility}), with $\chi_{\rm sc}(q,0)$ given by Eq.\
(\ref{chisc}), that the coefficient $\gamma_0(r_{\rm s})$ is fixed
by the compressibility via
\begin{equation} 
\frac{\kappa_0}{\kappa} = 1 - \frac{\sqrt{2}}{\pi} \gamma_0(r_{\rm s})
r_{\rm s}.
\end{equation} 
The local-field correction was determined in Ref.\ \cite{Iwamoto} only
at discrete values of $r_{\rm s}$.  We use a simple interpolating
procedure to obtain $G(q)$ for arbitrary values of $r_{\rm s}$ in the
entire interval $0 \leq r_{\rm s} \leq 40$.

At the value $r_{\rm s} = \bar{r}_{\rm s}$ where the inverse
compressibility changes sign, with $r_{\rm s} \approx 2.03$ according to
the Monte Carlo data \cite{TC}, the dielectric function becomes negative
for small wavevectors.  As for a 3DEG, the long-wavelength solution of
the condition (\ref{condition}), with a factor $q$ included instead of
$q^2$, changes here qualitatively.  The resulting equation is quadratic
in $q$ rather than $q^2$,
\begin{equation} \label{expand_2d}
a_0 + a_1 \hat{q} + a_2 \hat{q}^2 = 0,
\end{equation}  
with 
\begin{equation} 
a_0 = \frac{(\sqrt{2} /\pi) r_{\rm s}}{1 - (\sqrt{2} /\pi)
\gamma_0(r_{\rm s})r_{\rm s}},
\end{equation}  
while the two remaining coefficients are again best represented by their
numerical values.

\begin{figure}
\begin{center}
\psfrag{im}[t][t][.8][0]{Im[$\hat{q}(r_{\rm s})$]}
\psfrag{re}[t][t][.8][0]{Re[$\hat{q}(r_{\rm s})$]}
\psfrag{rs}[t][t][.8][0]{$r_s$}
\psfrag{q}[t][t][.8][0]{}
\psfrag{1}[t][t][.8][0]{$1$}
\psfrag{2}[t][t][.8][0]{$2$}
\psfrag{3}[t][t][.8][0]{$3$}
\psfrag{10}[t][t][.8][0]{$10$}
\psfrag{20}[t][t][.8][0]{$20$}
\psfrag{30}[t][t][.8][0]{$30$}
\psfrag{40}[t][t][.8][0]{$40$}
\psfrag{0.5}[t][t][.8][0]{$0.5$}
\psfrag{1.5}[t][t][.8][0]{$1.5$}
\psfrag{2.5}[t][t][.8][0]{$2.5$}
\includegraphics[width=8.cm]{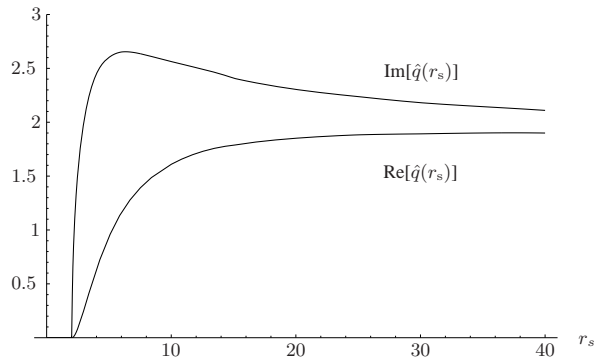}
\end{center}
\caption{Real (Re) and imaginary (Im) part of the wavevector
$\hat{q}(r_{\rm s})$ solving the condition (\ref{expand_2d}).
\label{fig:q_sol_2d}}
\end{figure}

For $r_{\rm s} < \bar{r}_{\rm s}$, no physical solution is found,
implying that in this entire regime screening is absent and the plasmon
mode gapless, as in the weak-coupling limit [see below Eq.\
(\ref{2dcon})].  For $r_{\rm s} > \bar{r}_{\rm s}$, a complex solution
with positive real and imaginary parts emerges, signalling an
exponentially damped CDW in a 2DEG (see Fig.\ \ref{fig:q_sol_2d}).
Contrary to the 3D case, the imaginary part of the solution is larger
than the real part in the entire regime where the parametrization
applies.

In conclusion, 2DEGs and 3DEGs at negative compressibility, where test
charges are overscreened, were argued to have an exponentially damped
CDW as ground state.  The wavevector characterizing this state is
complex and varies with the electron number density.  The real part
vanishes above the critical density, where the inverse compressibility
changes sign and the system becomes homogeneous.

\acknowledgments
I wish to thank B. Rosenstein for the kind hospitality at NCTU and
acknowledge helpful discussions with him and P.~Phillips.  This work was
funded by the National Science Council (NCS) of Taiwan, R.O.C.

\end{document}